\documentclass{llncs}
\usepackage{graphicx}

\begin{document}

\frontmatter          
\pagestyle{headings}  
\addtocmark{Pilot Blade Study} 
\mainmatter              

\title{Study of the CMS Phase-1 Pixel Pilot Blade Reconstruction}
\titlerunning{Study of the CMS Phase-1 Pixel Pilot Blade Reconstruction}

\author{Tamas Almos Vami \and Viktor Veszpremi
for the CMS Collaboration}
\authorrunning{Tamas Almos Vami and Viktor Veszpremi}
\tocauthor{Tamas Almos Vami and Viktor Veszpremi for the CMS Collaboration}
\institute{Wigner Research Centre for Physics, Budapest, Hungary \\ Email: \email{vami.tamas@wigner.mta.hu}}

\maketitle
\begin{abstract}
The Compact Muon Solenoid (CMS) detector is one of two general-purpose detectors that measure the products of high energy particle interactions in the Large Hadron Collider (LHC) at CERN. The silicon pixel detector is the innermost component of the CMS tracking system. The detector which was in operation between 2009 and 2016 has now been replaced with an upgraded one in the beginning of 2017. During the previous shutdown period of the LHC, a prototype readout system and a third disk was inserted into the old forward pixel detector with eight prototype blades constructed using the new digital read-out chips. Testing the performance of these pilot modules enabled us to gain operational experience with the upgraded detector. In this paper, the reconstruction and analysis of the data taken with the new modules are presented including information on the calibration of the reconstruction software. The hit finding efficiency and track-hit residual distributions are also shown.
\keywords{CMS, Semiconductor detector, Phase1 Upgrade}
\end{abstract}
\section{Introduction}
\vspace{-6pt}
The CMS pixel detector \cite{Phase0} was replaced in March 2017 with the "Phase-1 upgrade" \cite{Phase1} pixel detector. A prototype system, eight pilot blades (PB) with a new Phase-1 digital read-out scheme, was installed in 2014. The new blades are located in a third disk, next to the existing two disks of the forward pixel detector.

These modules enabled us to gain commissioning and operational experience with the Phase-1 detector. Studies verified the concepts of the new read-out and DAQ electronics, as well as, the data-reconstruction with the new sensors. Before data taking, Monte Carlo simulation was used to verify the software \cite{DP}, while the results of the data reconstruction process are presented in this paper \cite{DP2}.

\section{Results}
\vspace{-6pt}
\subsection{Cluster and hit positions}
\begin{figure}[!ht]
  \centerline{
  \includegraphics[width=.99\textwidth]{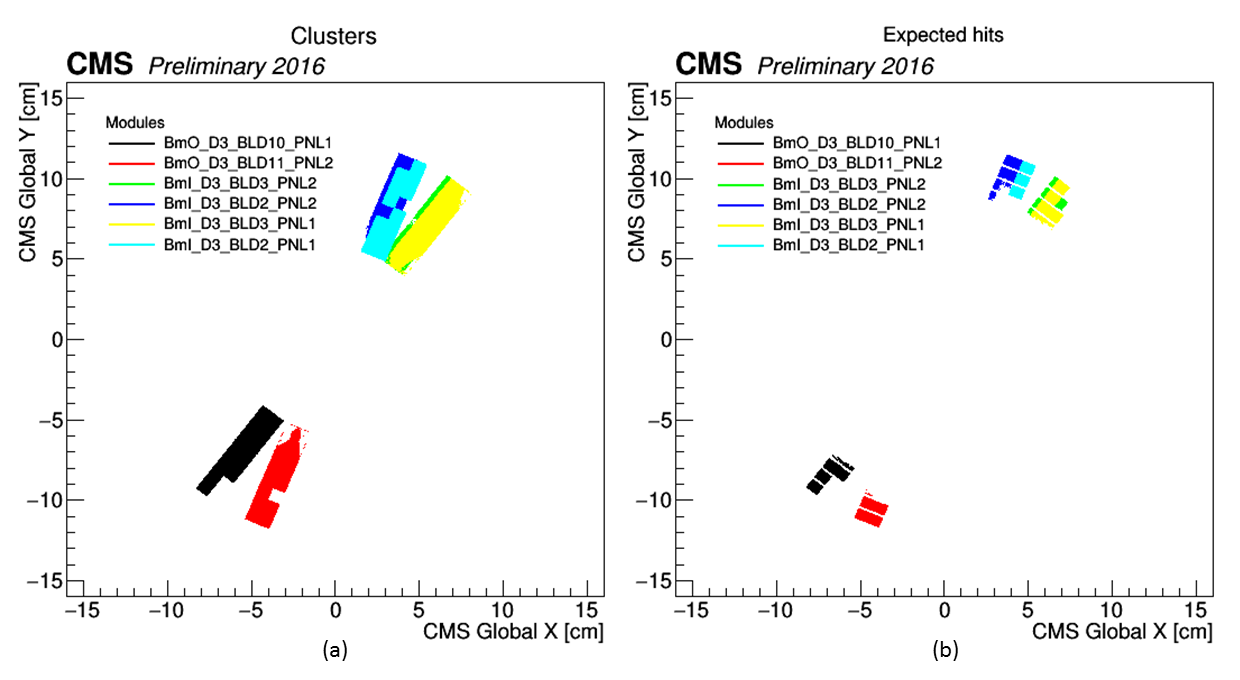}}
  \caption{(a) Cluster positions in each working pilot blade module and (b) expected hits in a transverse plane of the CMS coordinate system.}
  \label{fig:1}
\end{figure}

The pilot blade modules are made of 2$\times$8 Read-Out Chips (ROCs) arranged in rectangular shapes situated above the silicon sensors. Out of eight installed modules, six were successfully used in data taking. Hit pixels in the PB are read out in zero-suppression mode and get clusterized by the reconstruction software. The positions of the clusters are determined by their barycenter. Fig.\,\ref{fig:1}\,(a) shows the measured cluster positions projected into a transverse plane of the CMS coordinate system. The x-axis of the right-handed frame points from the geometric center of CMS to the center of the LHC ring, while the y-axis points upwards along a line that is perpendicular to x and the beam axis. 


Fig.\,\ref{fig:1}\,(b) shows the expected hits in the same view. Expected hits are trajectory states at the mid-plane of the PB sensors obtained by extrapolating collision tracks that are reconstructed in the forward pixel detector. The pilot blades are only spectators in tracking. They are located near the edge of the tracking coverage. In order to discard the fringes of ROCs, where the uncertainty in the track extrapolation is large, fiducial regions are defined. These are visible as rectangular shapes in Fig.\,\ref{fig:1}\,(b). Color coding identifies individual modules. The legend is to be understood in the following way: the pixel detector is divided into 4 quadrants by the $z=0$ and the $x=0$ planes, called BpI, BmI, BpO, and BmO. The letter p (m) means that the z coordinate of the module is positive (negative). The letter I (O) means that the x coordinate of the module is positive (negative). D3 means that the pilot system is installed on the support structure of a third disk. BLD shows the serial number of the blade counted in the transverse plane, while PNL describes which side (panel) of that blade is plotted.

\subsection{Timing and misalignment}
\begin{figure}[!ht]
  \centerline{
  \includegraphics[width=.99\textwidth]{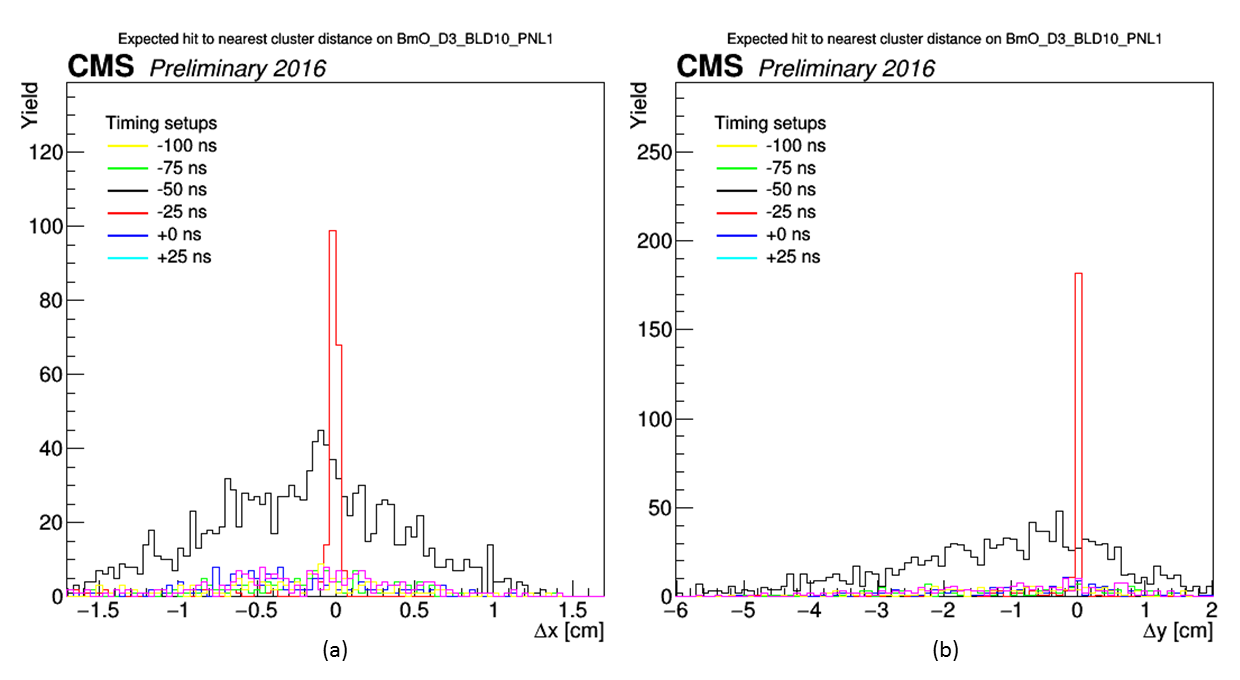}}
  \caption{Expected hit to nearest cluster distance in the (a) azimuthal and (b) radial direction. Red curve indicates the best setting for trigger latency.}
  \label{fig:2}
\end{figure}

The detector buffers hit-pixels in 25\,ns time windows matching the period of the LHC clock. Hits are retrieved upon receiving a read-out trigger signal generated by the real-time event-selection system of the CMS detector. First the correct trigger latency setting was found. Measurements were taken with several trigger latency assumptions, shown in different colors in Fig. \ref{fig:2}. The corresponding values stand for the adjustment of the trigger latency with respect to an initial estimate. The best setting is selected using the expected hit to nearest cluster distance distributions. The optimal latency is signaled by a narrow peak near zero indicating that the associated clusters truly belong to the particle tracks.

\begin{figure}[!ht]
  \centerline{
  \includegraphics[width=.99\textwidth]{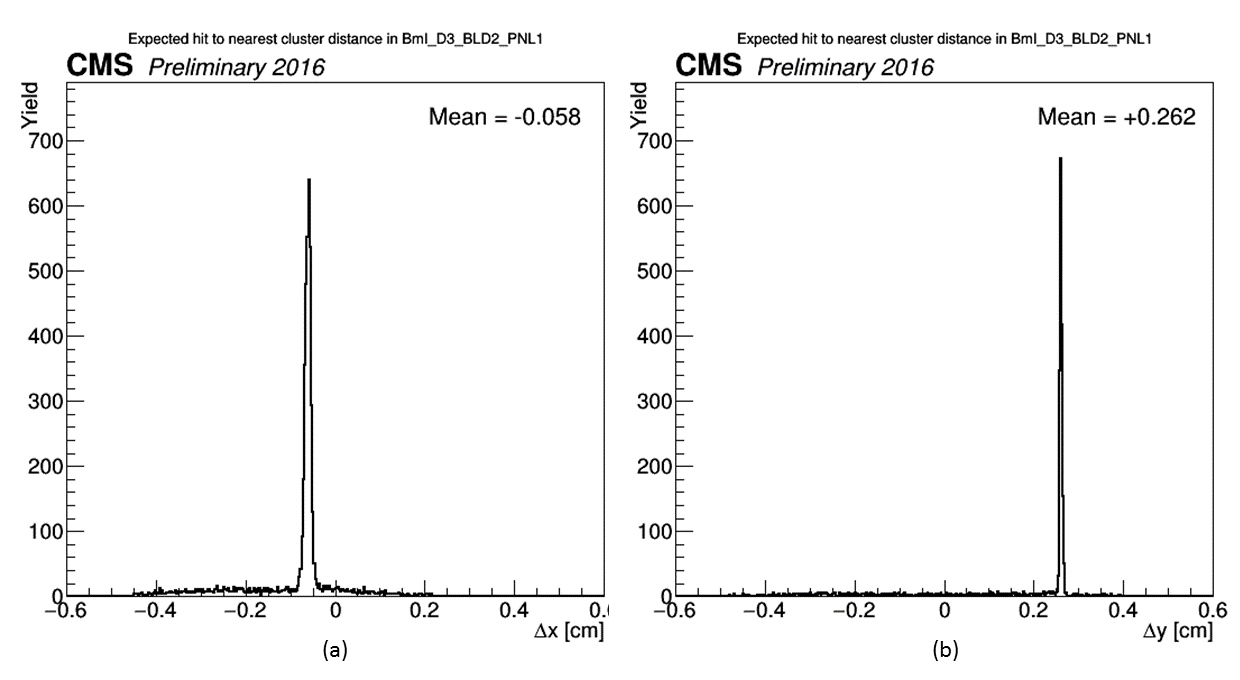}}
  \caption{Expected hit to nearest cluster distance demonstrating the misalignment of the module in the (a) azimuthal and (b) radial direction.}
  \label{fig:3}
\end{figure}

Module misalignments are seen as the mean of these distributions. For demonstrative purposes, measurements with a highly misaligned module are shown in Fig. \ref{fig:3}. The displacement is -0.058\,cm in the azimuthal and +0.262\,cm in the radial direction. Module positions were readjusted only in their local x-y plane by centering these distributions before computing the sensor efficiencies.

\subsection{Residual and efficiency}
\begin{figure}[!ht]
  \centerline{
  \includegraphics[width=1.2\textwidth]{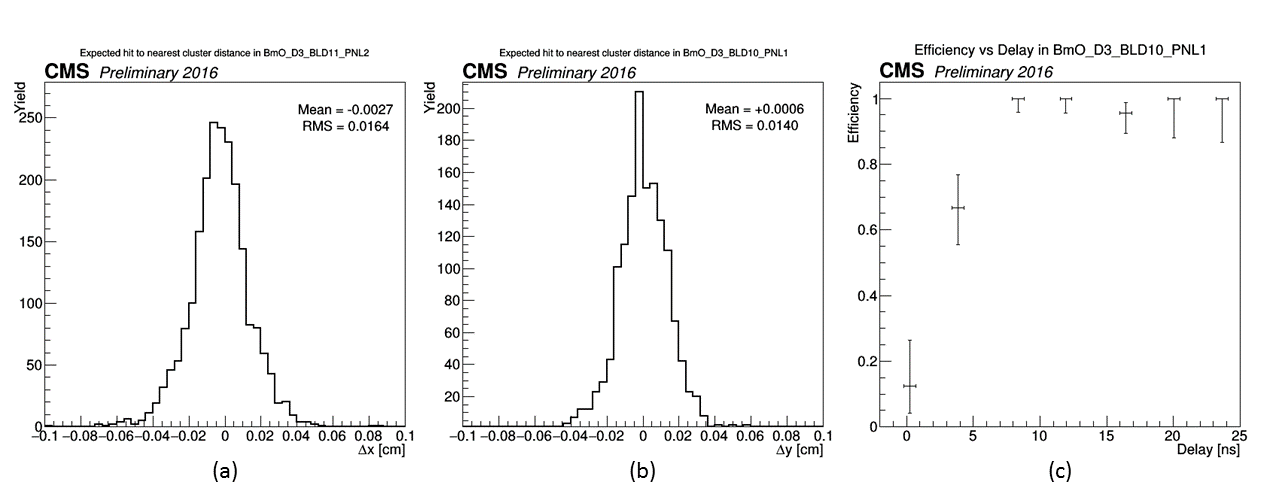}}
  \caption{Expected hit to nearest cluster distance in the azimuthal/local x (a) and the radial/local y direction (b) at timing settings when the ROC is fully efficient. The efficiency of one selected ROC on blade 10 vs the phase delay settings is shown (c). This ROC is located in the reliable tracking region, at the outer end of the module. Its misalignment is reasonable compared to the size of the cluster matching window.}
  \label{fig:4-1}
\end{figure}

After finding the trigger latency, before fine-tuning the timing, the ROCs return hit-pixel measurements belonging to the triggered collisions and to one of the adjacent ones. The next step is to adjust the phase at which the clock and trigger signals arrive to the ROCs in order to maximize the read-out efficiency. A scan of the clock phase delay was carried out in 4\,ns steps (Fig. \ref{fig:4-1} (c)). 

Efficiency is defined as the fraction of all the expected hits where a matching cluster is found within a 2\,mm radius. Using the optimized settings the accuracy of the cluster position measurement is within two pixels.

\section{Conclusions}
\vspace{-6pt}
A reconstruction analysis software was developed in order to help the commissioning and verification of the new read-out and DAQ electronics. Trigger latency and timing of the detector were determined. The new modules demonstrated full efficiency with the optimized settings.


\end{document}